# Electrical control of exchange spring in antiferromagnetic metals

*Yuyan Wang, Xiang Zhou, Cheng Song,* [a] *Yinuo Yan, Shiming Zhou, Guangyue Wang, Chao Chen, Fei Zeng, and Feng Pan*

Y. Y. Wang, Prof. C. Song, Y. N. Yan, G. Y. Wang, Dr. C. Chen, Dr. F. Zeng, Prof. F. Pan

Key Laboratory of Advanced Materials (MOE), School of Materials Science and Engineering, Tsinghua University, Beijing 100084, China.

X. Zhou, Prof. S. M. Zhou

Shanghai Key Laboratory of Special Artificial Microstructure and Pohl Institute of Solid State, Physics and School of Physics Science and Engineering, Tongji University, Shanghai 200092, China.

Antiferromagnet (AFM) spintronics have emerged as a fascinating research area and stimulated intense interest due to their potential for ultrafast and ultrahigh-density spintronics.[1–3] Much beyond the passive role of AFM as pinning layer, great efforts have been made in manipulating the AFM spins to realize the AFM-based memory resistors[3,5] and tunneling anisotropic magnetoresistance[6–8]. Experimental methods are successfully introduced including triggering partial rotation of AFM spins by neighbored FM,[6–8] reorientating spin axis of IrMn and FeRh through field cooling procedure,[3,5] and mediating metamagnetic transition temperature of FeRh using ferroelectricity in $BaTiO_3$ via strain transfer.[9,10] Besides, spin-transfer torques in AFM induced by spin-polarized current have been theoretically predicted,[11–13] lack of direct detection from experiments.

---

[a] E-mail: songcheng@mail.tsinghua.edu.cn

Controlling magnetism by means of electric fields manifests great superiority for the future development of low power spintronics, and favorable progress has been made in ferromagnetic (FM) systems.[14–18] The electric-field-assisted reversible switching in magnetic tunnel junctions enables a greatly reduced current density of ~$10^4$ A cm$^{-2}$, much smaller than that of normal spin-transfer torque switching (~$10^6$ A cm$^{-2}$),[19,20] offering advantages towards energy-efficient devices. In contrast to FM, the spin structures in AFM are hard to probe since they exhibit no net magnetic moment as a single layer. Recently, electrical control of interfacial coupling has been successfully demonstrated in FM/AFM heterostructures containing multiferroic $Cr_2O_3$[21] or $BiFeO_3$ as the dielectric layer,[22–25] through coupling between ferroelectricity and antiferromagnetism in the insulated AFM. However, for metallic AFM such as IrMn and FeMn, which plays an irreplaceable role in traditional spintronic devices,[1,26,27] direct electrical control remains challenging because of the screening effect by the surface charge. Generally, the manipulation is confined to a limited depth of atomic dimensions, which is insufficient to form a stable AFM exchange spring. Even for the recent report on the electric-field control of transformation between AFM and FM states in relatively thick FeRh film (22 nm), it is dominated by the strain effect from the $BaTiO_3$ substrate.[9]

In this work, we specially adopt an ionic liquid as the dielectric gate to modulate the exchange spring in IrMn AFM with different thicknesses. Compared with conventional solid gate insulator, the electric double layer (EDL) transistor has been developed as a powerful device structure allowing an extremely high electric field effect and penetrating a deeper thickness.[18,28–31] Fortunately, the exchange spring in AFM is intimately based on the thickness, making the modulation more explicit.[7,8] A direct observation of the adjustable exchange bias with gate voltages ($V_G$) is revealed by anomalous Hall effect (AHE) of

substrate/[Co/Pt]/IrMn. The $V_G$-controlled spin behaviors in AFM would provide an attractive alternative towards AFM spintronics.

The stack structures of Ta(4)/Pt(8)/[Co(0.5)/Pt(1)]$_4$/Co(0.5)/Pt(0.6)/IrMn($t$) (unit in nanometers) with different IrMn thicknesses ($t$) of 3 nm, 5 nm and 8 nm are deposited, capped by a 2 nm-thick HfO$_2$ (a high-permittivity material) to prevent the direct chemical reaction between metallic IrMn and ionic liquid. Then the multilayers were patterned into Hall devices, with a drop of ionic liquid utilized on top of HfO$_2$ as the electrolyte (**Figure 1**a). The ionic liquid film used here contains the following cation and anion: N,N-diethy1-N-(2-methoxyethy1)-N-methylammonium (DEME$^+$) and bis(trifluoromethylsulfony1)-imide (TFSI$^-$). By applying a gate voltage, the cations and anions move respectively to the channel and the gate electrode. An EDL is formed over the channel surface to accumulate electron carriers (Figure 1b).[18,28] The freezing point of this ionic liquid is around 210 K, below which the electric field in EDL is conserved at the solid state even after removing the external voltage.[30,31] Generally, the dielectric gate should be placed next to the modulated layer in order to generate a strong electric field effect.[28,30,31] Although the existence of HfO$_2$ in our experiment will weaken the electric field by adding an extra thickness of 2 nm to the original thickness of EDL,[32] the electric field effect generated here is still much larger than that of simply using traditional ~100 nm-thick dielectric layers.[18] As indicated in the insets of Figure 1b, the exchange spring formed in IrMn could be manipulated by the electric field and transfer the modulation to the [Co/Pt]/IrMn interface, in analogy to a real mechanical spring which transfers the force (inset of Figure 1b).

To verify the effect of $V_G$ on antiferromagnetic IrMn, we firstly investigate the AHE of the Hall devices, which reflects the magnetization properties of [Co/Pt]/IrMn at vertical fields ($H$), referring to the easy axis of Co/Pt.[33] In **Figure 2**a, the $V_G$-dependent Hall resistance ($R_{Hall}$) curves when $t$ = 3 nm are shown. The data were collected at 10 K after the field-cooling procedure, where a field of +10 kOe is applied to pin the interfacial spins and produce the exchange bias. A strong perpendicular magnetic anisotropy of the sample is obtained from the squared hysteresis loops in the inset of Figure 2a. Remarkably, the expanded scale of $R_{Hall}$-$H$ loops in Figure 2a present a close dependence of exchange bias on opposite sign of $V_G$. In the initial state as $V_G = 0$ V, a hysteresis loop featured with sharp reversal and clear bias towards negative fields is observed. Then we heat up the device to 300 K and apply a $V_G$ of −1.6 V for 30 min to sufficiently form a stable EDL on top of IrMn,[30] followed by a field cooling procedure and taking the measurements at 10 K. Surprisingly, the whole $R_{Hall}$-$H$ loop moves towards the negative $H$ when $V_G$ is −1.6 V, which indicates an increasing magnitude of bias field ($H_E$) compared with that at 0 V. Afterwards, we apply a $V_G$ of +1.8 V at 300 K in the same way, the value of which is higher than −1.6 V to fully polarize the ionic liquid. In contrast, the hysteresis loop moves backwards to positive $H$, weakening the initial exchange bias and even generating a positive $H_E$. To verify the reversible manipulation of exchange bias by gate voltages, successive measurements were carried out afterwards by applying the $V_G$ of −2 V and +2.2 V, generating the similar signals as that of −1.6 V and +1.8 V, respectively. The $H_E$ and $H_C$ extracted from the AHE curves varying with different $V_G$ are summarized in Figure 2b to make a clear comparison. It is found that the negative $V_G$ strengthens the negative exchange bias and increases the $H_C$, while the positive $V_G$ weakens the initial bias.

Given that the conductive metal generally has a screening effect on the electric field,[34] we now address the question whether the manipulation by gate voltage exists with increasing IrMn thickness. Figure 2c exhibits the $V_G$-dependent AHE effect at 10 K for $t = 5$ nm. A larger exchange bias is observed due to the stronger exchange coupling between 5 nm-thick IrMn and Co/Pt (inset of Figure 2c). Similar to that with 3 nm-thick IrMn, the negative $V_G$ also shifts the original $R$-$H$ loops towards the direction of negative $H$, strengthening the exchange bias, and the positive $V_G$ affects oppositely. Despite the similar tendency in Figure 2b and 2d, the modulation of $H_E$ and $H_C$ by $V_G$ in 5 nm-thick IrMn is much smaller than that in 3 nm-thick IrMn. When the thickness of IrMn increases to 8 nm, the electrical control of exchange bias does not work, even with $V_G$ up to 2.2 V. These thickness-dependent experiments clearly illustrate the electrical control of the exchange spring in AFM, which will be disscussed later.

We have explored the modulation of $H_E$ as a function of temperature to verify the electrical manipulation of AFM. **Figure 3**a depicts the temperature-dependent $H_E$ detected by AHE effect in the Hall device when $t$ = 3 nm, 5 nm and 8 nm. As expected, the blocking temperature ($T_b$) of IrMn raises as $t$ increases, which characterizes the disappearance of the effective exchange bias between Co/Pt and IrMn. The $T_b$ for 3 nm- and 5 nm-thick IrMn are no more than 50 K and 200 K respectively, while the exchange bias for $t$ = 8 nm could be retained to 300 K. Then we collect the $H_E$ at different $V_G$ and summarize the $\Delta H_E = H_E (V_G) - H_E (V_G = 0\text{ V})$ when $t$ = 3 nm and 5 nm in Figure 3b and 3c respectively. Clearly, a negative $V_G$ of −2.0 V will strengthen the exchange bias (positive $\Delta H_E$) towards negative direction, in contrast to the weakened bias at +2.2 V (negative $\Delta H_E$), which corresponds well to the results in Figure 2. Note that the electrical modulation of exchange bias only operates below $T_b$ for different $t$ when IrMn is antiferromagnetic. This indicates that the electrical tune indeed acts on AFM moments, meanwhile proving the repeatability of the experiments.

In order to explicit the mechanism behind the electrical modulation varying with IrMn thicknesses, we should note that although the exchange bias is usually considered as an interfacial effect between FM and AFM,[35] the interfacial AFM spins are coupled tightly to the exchange spring in AFM.[35,36] Remarkably, the change of $H_E$ and $H_C$ is probably ascribed to the electrical modulation of IrMn exchange spring, as discussed below. In general, the electric field-induced modification of metallic FM could be realized by changing the electron density at Fermi level $E_F$, which determines the intrinsic magnetic properties, such as the magnetic moment and magnetic anisotropy.[34] While for AFM, although the net magnetization is zero as a whole taking the antiparallel spins into account, the electron density at $E_F$ would affect the magnetic anisotropy of AFM, in analogy to the scenario of FM. The domain wall width $\delta_W$, in which an exchange-spring is formed, depends on the anisotropy constant of AFM $K_{AFM}$: $\delta_W = \pi\sqrt{A_{AFM}/K_{AFM}}/2$, where $A_{AFM}$ is the exchange stiffness.[7,36] Their close relationship is also convinced by the different static and dynamic behaviors of exchange spring in IrMn and FeMn, which possess different $K_{AFM}$.[36] It is then expected that the spin structure, especially the stability of exchange spring in IrMn is modified by $V_G$, probably due to the change of magnetic property of Mn in IrMn through injecting (negative $V_G$) and extracting electrons (positive $V_G$). To be specific, negative $V_G$ could increase the magnetic anisotropy of IrMn and stabilize the exchange spring, thus the interfacial interaction energy is increased to strengthen the pinning effect of IrMn. While for $t = 3$ nm, the gradual reversal in the descending branch (reversal from positive $R_{Hall}$ to negative $R_{Hall}$) at negative $V_G$ could be easily interpreted (Figure 2a), indicating the stabilizing of the interfacial spins in IrMn, which directly hinders the reversal process of Co/Pt moments. On the contrary, the positive $V_G$ would relax the negative pinning effect which is exerted by the field-cooling procedure, and the magnetization reversal is completed gradually in the ascending branch (Figure 2a).

We then attempt to make a deep insight into the intrinsic physics of the electrical manipulaton of metallic AFM. Inspired by the configuration of field-effect transistor which enables the direct detection of carrier densities, we have investigated the longitudinal resistance as a function of $H$ ($R_{AMR}$) of the Hall device substrate/IrMn(3, 5, 8 nm)/HfO$_2$ without Co/Pt FM, where IrMn is the only conductive medium. When $t = 3$ nm, the channel resistance as a function of time by applying a $V_G$ of −2 V for 30 min and then setting the $V_G$ to zero is recorded in **Figure 4**a. It is found that the resistance decreases sharply and reaches a nearly constant value after applying −2 V for 8 min. When the $V_G$ is set to zero, the effect of gate voltage still keeps although a slight resistance recovery is observed. Specifically, the created EDL under negative $V_G$ injects the free electrons into the bulk part of IrMn, thus adding the carriers which participate in the conduction. On the contrary, the gradual increase of the resistance when applying a $V_G$ of +2.2 V (Figure 4b) is due to the extraction of charge carriers which accumulate at the interface of IrMn/HfO$_2$ (Figure 1b). Note that the migration of ions in ionic liquid is a dynamic process, which needs some time (several minutes) to form a stable EDL under $V_G$.[30] Accordingly, the change of the resistance through injecting or extracting electrons, motivated by the electric field in EDL, also presents a gradual procedure. The influence of ionic current could be excluded since nearly no leakage current is detected at ±2.2 V, the highest voltage we used. A closer inspection of the curves shows a rapid process of resistance decrease at negative $V_G$ in contrast to the gradual procedure at positive $V_G$, possibly ascribed to different dynamic processes of injecting and extracting electrons. The $R_{AMR}$ in single-layered IrMn by sweeping $H$ under different voltages (0 V, −2 V, +2.2 V) are measured at 10 K, and different resistance states are clearly observed in Figure 4c, indicating the effective electrical control of conducting properties of IrMn. It is inferred that the change of the charge carriers in IrMn might affect the intrinsic electronic structure and magnetic moment of Mn, thus modulating the magnetic anisotropy of IrMn and the exchange bias, in analogy to the electrical tune of metallic FM. X-ray photoelectron spectroscopy (XPS) was adopted to analyze the chemical states of Mn in substrate/IrMn/HfO$_2$. As shown in Figure 4d, similar peak position and intensity of Mn 2$p$ under different $V_G$ verify that the valence state is unchanged. We also take the measurements for the sample without HfO$_2$ as a comparison, where the ionic liquid directly contacts with metallic IrMn. The greatly reduced Mn signals imply the reaction or corrosion between Mn and ionic liquid, confirming the indispensable role of HfO$_2$ as an insulating layer.

It is generally accepted that the electric field could not penetrate into the bulk of traditional metals and is confined to a depth on the order of atomic dimensions, due to the screening effect.[34] However, it is not surprising that the exchange bias could still be changed by different $V_G$ when $t$ is up to 5 nm (Figure 2c), considering the following two aspects. On one hand, substantial electric-induced effects can be realized by using the EDL though injecting or extracting electrons from the film, and act on a deeper thickness.[30] This is evidenced by the modification of magnetism in 4 nm-thick FePt,[34] in contrast to a thickness of 1~2 atomic layer in La$_{0.8}$Sr$_{0.2}$MO$_3$ modulated by ferroelectric Pb(Zr$_{0.2}$Ti$_{0.8}$)O$_3$.[37] Also, the higher resistivity of IrMn than traditional metal also induces wider modulation depth. On the other hand, the tunable exchange bias in 5 nm-thick IrMn further confirms the electrical modulation of exchange spring, since the electric-field effect is hard to reach the interface between Co/Pt and IrMn directly. As $t$ increases from 3 nm to 5 nm, a more rigid exchange spring is created to stabilize the AFM moments as well as enhance the original exchange coupling ($V_G$ = 0 V).[7] In this case, the electrical modulation of exchange spring becomes harder and the shift of exchange bias is less. This is evidenced by the reduced electrical control of channel resistance in 5 nm-thick IrMn (Figure 4e), since the conduction is contributed by the whole IrMn while the electric field only modulates part of IrMn. When it reaches 8 nm, comparable to the domain-wall width of IrMn (~7.8 nm), which is the upper limit for the formation of a complete exchange spring,[7,8] the manipulation of IrMn by electric field is powerless and the interfacial spin behaviors are unchanged. As expected, the resistance of 8 nm-thick IrMn could hardly be manipulated by $V_G$ (Figure 4e), consistent with the unchangeable exchange bias in [Co/Pt]/IrMn(8 nm)/HfO$_2$. On the whole, the change of carrier densities and modulation of electrons structure might have a profound influence on the magnetic anisotropy of IrMn, which directly affect the exchange spring and changes the interfacial pinning effect.

Up to now, we have presented the manipulation of exchange bias by different $V_G$ and proposed the mechanism as controlling the behaviors of exchange spring in IrMn. In order to directly detect the induced change of interfacial uncompensated spins in IrMn, a FM/AFM system consists of insulating Y$_3$Fe$_3$O$_{12}$ (YIG) FM and IrMn AFM is adopted to investigate the $V_G$ control of exchange bias, where the electrical transport is purely originated from metallic

IrMn.[38,39] For this control experiment, a 20 nm-thick YIG film was deposited on $Gd_3Ga_5O_{12}$ (GGG) substrate by pulsed laser deposition, followed by a IrMn layer (3 nm) grown by magnetron sputtering without breaking the vacuum. The Hall resistance by sweeping in-plane *H* which is perpendicular to the current *I*, so-called planar Hall effect (PHE), are displayed in the inset of **Figure 5**a. It is well known that the PHE has been emerged as an effective method of investigating the magnetization reversal of ferromagnetic film and exchange bias system.[33,40] For a single IrMn layer deposited on $Si/SiO_2$ substrate, no recognizable PHE signals are observed, since IrMn exhibits zero net magnetization and the moments are hardly rotated by the external field. When coupled with ferromagnetic YIG processing in-plane easy axis of magnetization, the PHE curves of the IrMn layer (Figure 5a) enable us to experimentally probe the interfacial uncompensated spins due to the magnetic proximity effect between YIG and IrMn.[38]

Corresponding magnetization loop of YIG/IrMn bilayer is depicted in Figure 5b. Despite the unresolvable ferromagnetic hysteresis signals of YIG in the raw data (inset of Figure 5b), which are totally covered by the strong paramagnetic signals arising from GGG substrate at low temperature,[38] the loop in Figure 5b after subtracting the paramagnetic background clearly shows the magnetization reversal of YIG pinned by 3 nm-thick IrMn. Compared with the coercivity and bias field of the magnetization loop, their counterparts are comparable but slightly larger in the PHE loop, which reflects the magnetization reversal of the uncompensated moment in IrMn. It is not surprising because the interfacial spins of IrMn are motivated by YIG via the proximity effect and accomplish the reversal gradually with richer interface.[38] Thus, the comparison between PHE curves arising from IrMn and magnetization loop relevant to YIG confirms the coupling between them. An expanded scale of ±0.5 kOe is depicted in Figure 5 to see the details of the PHE at different $V_G$. As expected, the negative $V_G$ strengthens the bias effect by enhancing the interfacial uncompensated spins in IrMn, in contrast to the weakened exchange bias at positive $V_G$. Therefore, the electrical control of PHE here crucially demonstrate the change of interfacial spin behaviors as a result of the electrical modulation of AFM exchange spring, supporting the main idea of this work.

In summary, we have reported on the electric-field manipulation of metallic antiferromagnet IrMn using an ionic liquid as the gate electrode. From the anomalous Hall effect of the substrate/[Co/Pt]/IrMn devices, a reversible modulation of the exchange bias is observed at different $V_G$. Distinct controlling effects varying with IrMn thicknesses and operating temperatures are presented, proposing the manipulation of exchange spring in IrMn by electric field. In addition, electrical control of anisotropic magnetoresistance in single layered IrMn devices further reveals that the modulation of AFM exchange spring is correlated to the change of magnetic anisotropy as well as carrier density by injecting or extracting electrons. Admittedly, further theoretical calculations are required to explore the intrinsic mechanisms. This work provides a new approach towards modulating the exchange spring in metallic AFM via electrical methods, which should be significant in advancing the development of low-power-consumption AFM spintronics.

**Experimental Section**

*Sample Preparation*: The stack structures of Ta(4)/Pt(8)/[Co(0.5)/Pt(1)]$_4$/Co(0.5)/Pt(0.6)/IrMn(*t*) (unit in nanometers) with different IrMn thicknesses of 3 nm, 5 nm and 8 nm are deposited on the Si/SiO$_2$ substrate by magnetron sputtering, capped by a 2 nm-thick HfO$_2$ grown by ultra-high vacuum evaporation to prevent the direct chemical reaction between metallic IrMn and ionic liquid. Then the multilayers were patterned into Hall devices with channel width of 30 μm, using photolithography and ion milling techniques. A drop of ionic liquid (DEME-TFSI) was utilized on top of HfO$_2$ as the electrolyte. The ionic liquid covers the Hall channel of the devices, as indicated in Figure 1a. Due to the surface tension and viscosity, a liquid film is spread with thinner edge, where the average thickness is estimated to be ~5 μm.

*Magnetic Transport Measurement*: The Hall resistance and anisotropic magnetoresistance of the devices were measured by transverse $R_{\text{Hall}}$ and longitudinal $R_{\text{AMR}}$ respectively, using a physical property measurement system (PPMS), while a constant in-plane current of 500 μA was applied in the channel.

**Acknowledgments**

This work was supported by the National Natural Science Foundation of China (Grant Nos. 51322101, 51202125 and 51231004) and the National High Technology Research and Development Program of China (Grant no. 2014AA032904 and 2014AA032901).

Received:
Revised:
Published online:

[1] A. H. MacDonald, M. Tsoi, *Phil. Trans. R. Soc. A* **2011**, 369, 3098

[2] J. Sinova, I. Žutić, *Nat. Mater.* **2012**, 11, 368

[3] X. Martí, I. Fina, C. Frontera, J. Liu, P. Wadley, Q. He, R. J. Paull, J. D. Clarkson, J. Kudrnovsky, I. Turek, J. Kuneš, D, Yi, J-H. Chu, C. T. Nelson, L. You, E. Arenholz, S. Salahuddin, J. Fontcuberta, T. Jungwirth, R. Ramesh, *Nat. Mater.* **2014**, *13*, 367.

[4] X. Hu, *Adv. Mater.* **2012**, *24*, 294.

[5] D. Petti, E. Albisetti, H. Reichlová, J. Gazquez, M. Varela, M. Molina-Ruiz, A. F. Lopeandía, K. Olejník, V. Novák, I. Fina, B. Dkhil, J. Hayakawa, X. Martí, J. Wunderlich, T. Jungwirth, R. Bertacco, *Appl. Phys. Lett.* **2013**, *102*, 192404.

[6] B. G. Park, J. Wunderlich, X. Martí, V. Holy, Y. Kurosaki, M. Yamada, H. Yamamoto, A. Nishide, J. Hayakawa, H. Takahashi, A. B. Shick, T. Jungwirth, *Nat. Mater.* **2011**, *10*, 347.

[7] Y. Y. Wang, C. Song, B. Cui, G. Y. Wang, F. Zeng, F. Pan, *Phys. Rev. Lett.* **2012**, *109*, 137201.

[8] Y. Y. Wang, C. Song, G. Y. Wang, J. H. Miao, F. Zeng, F. Pan, *Adv. Funct. Mater.* **2014**, *24*, 6806.

[9] R. O. Cherifi, V. Ivanovskaya, L. C. Phillips, A. Zobelli, I. C. Infante, E. Jacquet, V. Garcia, S. Fusil, P. R. Briddon, N. Guiblin, A. Mougin, A. A. Ünal, F. Kronast, S. Valencia, B. Dkhil, A. Barthélémy, M. Bibes, *Nat. Mater.* **2014**, *13*, 345.

[10] I. Suzuki, M. Itoh, T. Taniyama, *Appl. Phys. Lett.* **2014**, *104*, 022401.

[11] Z. Wei, A. Sharma, A. S. Nunez, P. M. Haney, R. A. Duine, J. Bass, A. H. MacDonald, M. Tsoi, *Phys. Rev. Lett.* **2007**, *98*, 116603.

[12] Y. Xu, S. Wang, K. Xia, *Phys. Rev. Lett.* **2008**, *100*, 226602.

[13] R. Cheng, J. Xiao, Q. Niu, A. Brataas, *Phys. Rev. Lett.* **2014**, *113*, 057601.

[14] T. Maruyama, Y. Shiota, T. Nozaki, K. Ohta, N. Toda, M. Mizuguchi, A. A. Tulapurkar, T. Shinjo, M. Shiraishi, Y. Ando, Y. Suzuki, *Nature Nanotech*. **2009**, 4, 158.

[15] P. Li, A. Chen, D. Li, Y. Zhao, S. Zhang, L. Yang, Y. Liu, M. Zhu, H. Zhang, X. Han, *Adv. Mater.* **2014**, *26*, 4320.

[16] N. Lei, T. Devolder, G. Agnus, P. Aubert, L. Daniel, J.V. Kim, W. Zhao, T. Trypiniotis, R. P. Cowburn, C. Chappert, D. Ravelosona, P. Lecoeur, *Nat. Commun.* **2012**, *4*, 1378.

[17] H. Ohno, D. Chiba, F. Matsukura, T. Omiya, E. Abe, T. Dietl, Y. Ohno, K. Ohtani, *Nature* **2000**, *408*, 944.

[18] D. Chiba, T. Ono, *J. Phys. D: Appl. Phys.* **2013**, *46*, 213001.

[19] W. G. Wang, M. Li, S. Hageman, C. L. Chien, *Nat. Mater.* **2012**, *11*, 64.

[20] W. G. Wang, C. L. Chien, *J. Phys. D: Appl. Phys.* **2013**, *46*, 074004.

[21] W. Echtenkamp, Ch. Binek, *Phys. Rev. Lett.* **2013**, *111*, 187204.

[22] Y. H. Chu, L. W. Martin, M. B. Holcomb, M. Gajek, S. J. Han, Q. He, N. Balke, C. H. Yang, D. Lee, W. Hu, Q. Zhan, P. L. Yang, A. F. Rodríguez, A. Scholl, S. X. Wang, R. Ramesh, *Nat. Mater.* **2008**, *7*, 478.

[23] J. T. Heron, M. Trassin, K. Ashraf, M. Gajek, Q. He, S. Y. Yang, D. E. Nikonov, Y. H.

Chu, S. Salahuddin, R. Ramesh, *Phys. Rev. Lett.* **2011**, *107*, 217202.

[24] D. Yi, J. Liu, S. Okamoto, S. Jagannatha, Y. C. Chen, P. Yu, Y. H. Chu, E. Arenholz, R. Ramesh, *Phys. Rev. Lett.* **2013**, *111*, 127601.

[25] S. M. Wu, Shane A. Cybart, D. Yi, James M. Parker, R. Ramesh, R. C. Dynes, *Phys. Rev. Lett.* **2013**, *110*, 067202.

[26] A. Kohn, A. Kovács, R. Fan, G. J. Mclntyre, R. C. C. Ward, J. P. Goff, *Sci. Rep.* **2013**, *3*, 2412.

[27] V. M. T. S. Barthem, C. V. Colin, H. Mayaffre, M. H. Julien, D. Givord, *Nat. Commun.* **2013**, *4*, 2892.

[28] Y. Yamada, K. Ueno, T. Fukumura, H. T. Yuan, H. Shimotani, Y. Iwasa, L. Gu, S. Tsukimoto, Y. Ikuhara, M. Kawasaki, *Science* **2011**, *332*, 1065.

[29] L. H. Diez, A. B. Mantel, L. Vila, P. Warin, A. Marty, S. Ono, D. Givord, L. Ranno, *Appl. Phys. Lett.* **2014**, *104*, 082413.

[30] B. Cui, C. Song, G. Y. Wang, Y. N. Yan, J. J. Peng, J. H. Miao, H. J. Mao, F. Li, C. Chen, F. Zeng, F. Pan, *Adv. Funct. Mater.* **2014**, *24*, 7233.

[31] B. Cui, C. Song, G. A. Gehring, F. Li, G. Y. Wang, C. Chen, J. J. Peng, H. J. Mao, F. Zeng, F. Pan, *Adv. Funct. Mater.* **2015**, *25*, 864.

[32] P. Gallagher, M. Lee, J. R. Williams, D. G. Gordon, *Nat. Phys.* **2014**, *10*, 748.

[33] Y. Y. Wang, C. Song, F. Zeng, F. Pan, *J. Phys. D: Appl. Phys.* **2013**, *46*, 445001.

[34] M. Weisheit, S. Fähler, A. Marty, Y. Souche, C. Poinsignon, D. Givord, *Science* **2007**, *315*, 349.

[35] J. Nogués, I. K. Schuller, *J. Magn. Magn. Mater.* **1999**, *192*, 203.

[36] Y. Y. Wang, C. Song, G. Y. Wang, F. Zeng, F. Pan, *New J. Phys.* **2014**, *16*, 123032.

[37] C. A. F. Vaz, J. Hoffman, Y. Segal, J. W. Reiner, R. D. Grober, Z. Zhang, C. H. Ahn, F. J. Walker, *Phys. Rev. Lett.* **2010**, *104*, 127202.

[38] X. Zhou, L. Ma, Z. Shi, W. J. Fan, R. F. L. Evans, R. W. Chantrell, S. Mangin, H. W.

Zhang, S. M. Zhou, *Sci. Rep.* DOI: 10.1038/srep09183.

[39] J. H. Han, C. Song, S. Gao, Y. Y. Wang, C. Chen, F. Pan, *ACS Nano* **2014**, *8*, 10043.

[40] B. Sinha, T. Q. Hung, T. S. Ramulu, S. Oh, K. Kim. D. Y. Kim, F. Terki, C. Kim, *J. Appl. Phys.* **2013**, *113*, 063903.

**Figures and Captions:**

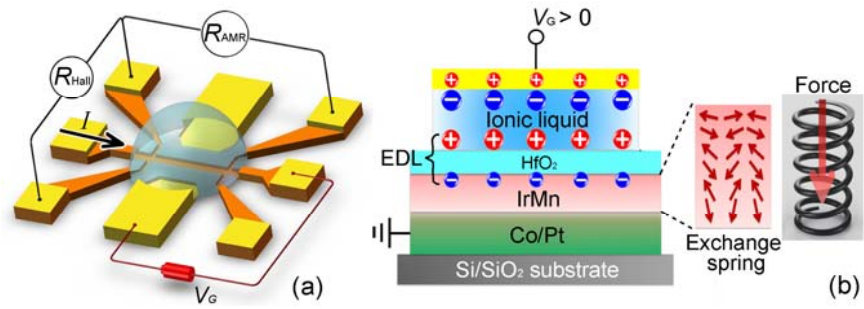

**Figure 1.** (a) Measurement configuration of the Hall devices with an ionic liquid electrolyte. (b) The schematic cross-section view along the channel to the gate electrode with positive $V_G$, and the charge distribution under the effect of electric double layer. The insets are the schematic of spins in IrMn exchange spring, and a illustration of mechanical spring which enables the transfer of the force from top to the bottom.

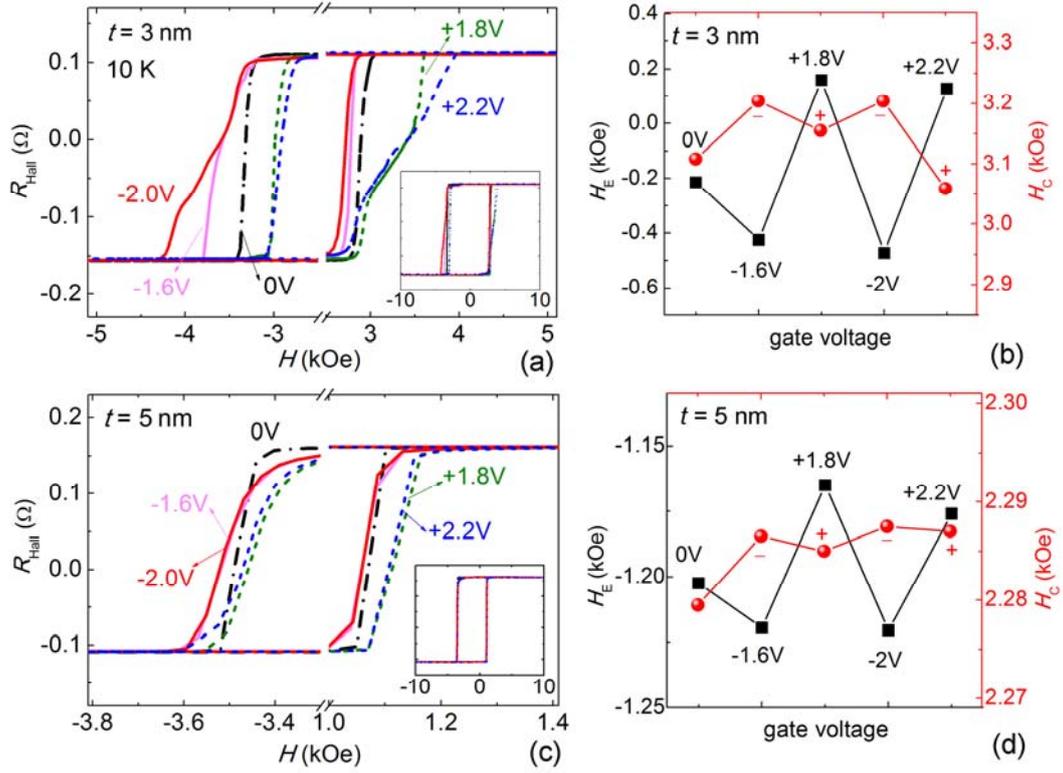

**Figure 2.** $V_G$-dependent Hall resistance $R_{Hall}$ acquired by sweeping vertical $H$ when (a) $t$ = 3 nm, (c) $t$ = 5 nm. The insets are the whole curves in the range of ±10 kOe. The $H_E$ and $H_C$ extracted from the AHE curves varying with different $V_G$ are summarized in (b) $t$ = 3 nm and (d) $t$ = 5 nm.

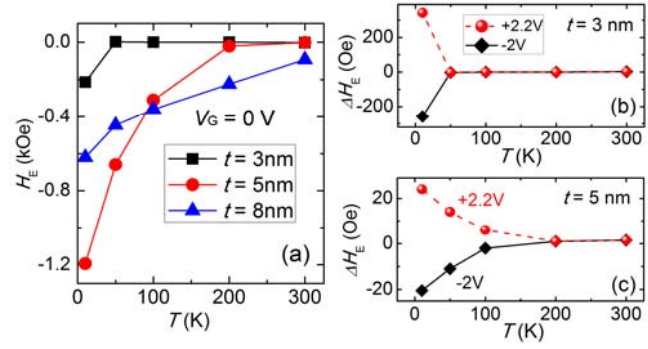

**Figure 3.** (a) Temperature-dependent $H_E$ extracted from the AHE curves with different $t$ when $V_G = 0$ V. (b) and (c) shows the temperature-dependent $\Delta H_E = H_E (V_G) - H_E (V_G = 0$ V$)$ when $t = 3$ nm and 5 nm respectively.

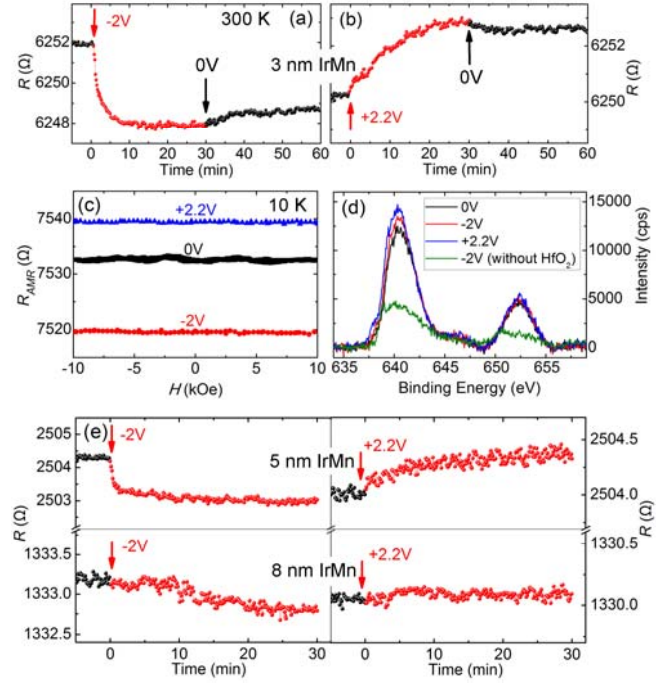

**Figure 4.** Channel resistance of the substrate/IrMn($t$ = 3 nm)/HfO$_2$ Hall device at 300 K as a function of time on applying a gate voltage for 30 min with (a) $V_G$ = −2 V, (b) $V_G$ = +2.2 V, and then setting the gate voltage to zero. (c) $R_{AMR}$ acquired at 10 K by sweeping vertical $H$ with different $V_G$. (d) Mn 2$p$ XPS spectra of samples with different $V_G$. (e) Channel resistance as a function of time on applying a gate voltage for 30 min when $t$ = 5 nm and 8 nm.

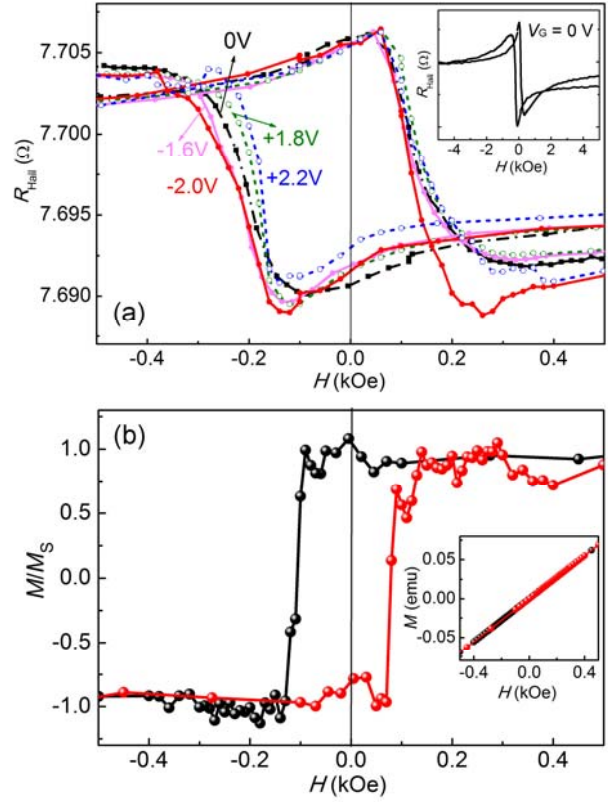

**Figure 5.** (a) $V_G$-dependent planar Hall resistance ($R_{Hall}$) of the sample GGG substrate/YIG (20 nm)/IrMn (3 nm) acquired by sweeping in-plane $H$ at 10 K. The inset is the PHE curve at 0 V in the scale of ±5 kOe. (b) Normalized magnetization loop of YIG/IrMn bilayers at 10 K after subtracting the paramagnetic background of GGG. The inset shows the raw data.